\documentclass[12pt]{article}
\usepackage{graphicx}
\usepackage{amsmath}

\begin{document}

\title{\textbf{OPTION PRICING AND HEDGING WITH TEMPORAL CORRELATIONS}}
\author{Lorenzo Cornalba$^{1}$, Jean--Philippe Bouchaud$^{2,3}$ and Marc Potters$%
^{3} $}
\date{\small $^{1}$\textit{\ }Laboratoire de Physique Th\'{e}orique de l'\'{E}cole
Normale Sup\'{e}rieure\\
24 rue Lhomond, 75231 Paris Cedex 5, France \\
$^{2}$ Service de Physique de l'\'{E}tat Condens\'{e}, Centre d'\'{e}tudes
de Saclay\\
Orme des Merisiers, 91191 Gif--sur--Yvette Cedex, France \\
$^{3}$ Science \& Finance\\
The Research division of Capital Fund Management\\
109--111 rue Victor-Hugo, 92532 Levallois Cedex, France \\
\vspace{0.5cm}
\today}
\maketitle

\begin{abstract}
We consider the problem of option pricing and hedging when stock returns are
correlated in time. Within a quadratic--risk minimisation scheme, we obtain a general formula, 
valid for weakly correlated non--Gaussian processes. We show that for Gaussian price increments, the
correlations are irrelevant, and the Black--Scholes formula holds with the
volatility of the price increments on the scale of the re--hedging. For
non--Gaussian processes, further non trivial corrections to the `smile' are
brought about by the correlations, even when the hedge is the Black--Scholes 
$\Delta $--hedge. We introduce a compact notation which eases the computations and
could be of use to deal with more complicated models.
\end{abstract}

\section{Introduction}

The assumptions underlying the Black--Scholes model \cite{Hull,Wilmott} --
that the dynamics of financial assets can be modelled by a continuous time
Gaussian process -- are very far from reality. It is well known, in
particular, that the empirical distribution of returns exhibits
`fat--tails'. The presence of these extreme events not only induces a
`smile' in the implied volatility of the options, but also destroys the
essence of the Black--Scholes pricing procedure: the possibility of
constructing a riskless hedge which replicates perfectly the option pay--off 
\cite{BS2,Sch,us}. A more subtle further effect, completely discarded by the
standard Black--Scholes model, is the presence of small, but significant,
temporal \textit{anticorrelations} in the series of stock returns.
Correspondingly, the variance of the $N$--day distribution of returns is
smaller than $N$ times the variance of the daily distribution (see \textit{%
e.g.} \cite{Lo}). On other financial assets, such as stock indices, one observes the
opposite effect of weak positive correlations. The order of magnitude of these
correlations or anticorrelations is $10\%$ from one day to the next. These
correlations lead, in principle, to the possibility of statistical arbitrage. 
However, one should also take into account transaction costs which significantly 
reduce the possibility of using these correlation in practice \cite{us}. In other words,
the presence of transaction costs allow the existence of non zero correlations,
which can be significant and therefore must be dealt with consistently in the
context of option pricing.

What is the influence of these correlations on the price of an option and on
the corresponding hedging strategy? The aim of this paper is to provide a
general answer to this question within a quadratic risk minimization sheme, 
in the case where these correlations are
small \footnote{%
The general correlated Gaussian case was considered in \cite{BS2}, and leads
to rather complex formulae for which no simple interpretation was given.}
(which is, in most cases of interest, amply justified), with only very few
assumptions on the actual distribution of returns. We calculate the first
order corrections both to the price and optimal hedge. In the Gaussian case,
we find that the effect of correlations can be compensated by a change in
the hedging strategy and therefore options should be priced using the
standard uncorrelated Black--Scholes model \footnote{This
can be easily seen within a Black--Scholes framework: see Appendix.}. The correct volatility to be
used is the one measured on the time scale of the rehedging, and not the one
corresponding to the terminal distribution -- \textit{i.e.} corresponding to
the maturity of the option. In other words, the `risk--neutral' measure that
one should use to price derivatives is simply an uncorrelated Gaussian
measure with the same elementary variance. These conclusions are not valid
if one considers a process for which the underlying distribution is
non--Gaussian. One then finds contributions to the option prices which
depend in general on the strength of the correlations, and on the higher moments of the
distribution (in particular on the kurtosis). The corrections to the option
price also depend on the past history of the price--increments of the
underlying. The detailed analysis of a specific example shows, in
particular, that anticorrelations change the effective kurtosis to be used
when computing `smile' corrections to the options prices. In general, the
correct kurtosis to be used is neither the local nor the global one, but
some effective kurtosis which depends on the strength of the correlations.

\section{\label{GS}General setting}

Our starting point is the global wealth balance approach discussed in \cite
{BS2,us}. We set, for simplicity, the interest rate to zero, and consider a
discrete time model with time interval $\tau $, within which no rehedging
takes place. The global wealth balance at time $T=N\tau $ for the writer of
the option reads 
\begin{equation*}
W=\mathcal{C}+H-\Theta ,
\end{equation*}
where $\mathcal{C}$ is the option premium, $\Theta $ is the pay--off of the
option, and $H$ is the result of the hedging strategy 
\begin{equation}
H=\sum_{n=0}^{N-1}\phi _{n}\Delta _{n+1}.  \label{eq1}
\end{equation}
In equation (\label{eq100}\ref{eq1}), $\Delta _{n+1}=x_{n+1}-x_{n}$ denotes
the change of the price $x$ of the underlying asset between time $n$ and $%
n+1 $, and $\phi _{n}$ is the amount of underlying in the portfolio at time $%
n$ \footnote{%
In the following, we assume for simplicity that the interest rate is zero.
It is easy to generalize the formulae obtained here when this rate is
non-zero: see \cite{us} for details.}. 

In incomplete markets, the price of an option is not unique since a risk--premium 
should be added. However, a convenient framework is the risk minimisation
scheme proposed in \cite{BS2,Sch,us}, that fixes both the price and
strategy $\mathcal{C}$, $\phi _{n}$ so as to minimize the total risk $\mathcal{R%
}$ associated to option writing, which we simply define as 
\begin{equation*}
\mathcal{R}^{2}=\left\langle W^{2}\right\rangle ,
\end{equation*}
where $\langle ...\rangle $ is an average over the objective probability at
time $0$. Minimizing with respect to $\mathcal{C}$ fixes the price of the
call option to be 
\begin{equation}
\mathcal{C}=\left\langle \Theta -H\right\rangle .  \label{eq200}
\end{equation}
If, on the other hand, we minimize $\mathcal{R}^{2}$ with respect to $\phi
_{n}$, we obtain an equation implicitly determining the optimal strategy $%
\phi _{n}^{\star }$, equation that reads, for $0\leq n<N$ 
\begin{equation}
\langle \Delta _{n+1}\left( \Theta -H\right) \rangle _{n}=\mathcal{C}\langle
\Delta _{n+1}\rangle _{n}  \label{general}
\end{equation}
The notation $\langle ...\rangle _{n}$ in the above equation means that the
average (again over the objective probability) is performed at time $n$ -- 
\textit{i.e.} with all the information available at this instant of time. 
This notation turns out to be very powerful, and could be used to handle 
very general situations. We
now analyze the above general and compact equations (\ref{eq200}) and (\ref
{general}) in several simple but important limiting cases.

\section{\label{zeroD}The case of zero conditional drift}

Let us define the conditional drift $\mu _{n}$ and the conditional variance $%
D_{n}$ at time $n$ as 
\begin{eqnarray*}
\mu _{n} &=&\left\langle \Delta _{n+1}\right\rangle _{n} \\
D_{n} &=&\left\langle (\Delta _{n+1}-\mu _{n})^{2}\right\rangle _{n}.
\end{eqnarray*}
Note that these quantities are the expected drift and variance for the next
time interval, and depend in general on the past (realized) history of price
changes.

We start by assuming that the drift $\mu _{n}$ is identically zero (this is
to say, the price process is a `martingale'). Then, since $\phi _{n}$ is
known at time $n$, one has that 
\begin{equation*}
\left\langle \phi _{n}\Delta _{n+1}\right\rangle =\left\langle \phi
_{n}\langle \Delta _{n+1}\rangle _{n}\right\rangle =0,
\end{equation*}
and therefore that $\left\langle H\right\rangle =0$. This implies that the
option price is nothing but the expectation value of the pay--off 
\begin{equation*}
\mathcal{C}=\left\langle \Theta \right\rangle .
\end{equation*}
To determine the optimal hedge, we note that, in general, one has 
\begin{eqnarray}
\left\langle \Delta _{n+1}H\right\rangle _{n}
&=&\sum_{m=0}^{N-1}\left\langle \phi _{m}\Delta _{m+1}\Delta
_{n+1}\right\rangle _{n}  \label{corr} \\
&=&\phi _{n}\left\langle \Delta _{n+1}^{2}\right\rangle
_{n}+\sum_{m=0}^{n-1}\phi _{m}\Delta _{m+1}\mu
_{n}+\sum_{m=n+1}^{N-1}\left\langle \phi _{m}\mu _{m}\Delta
_{n+1}\right\rangle _{n}.  \notag
\end{eqnarray}
In the case of vanishing drift $\mu _{n}=0$ we therefore have that 
\begin{equation*}
\left\langle \Delta _{n+1}H\right\rangle _{n}=\phi _{n}D_{n}
\end{equation*}
so that the optimal hedging strategy can be computed, from equation (\ref
{general}), as \cite{BS2,us} 
\begin{equation*}
\phi _{n}^{\star }=\frac{1}{D_{n}}\left\langle \Delta _{n+1}\Theta
\right\rangle _{n}.
\end{equation*}
\qquad

\section{\label{pert}Perturbative expansion for small drift}

Let us now turn to the case where the conditional bias $\mu _{n}$ is
non--zero but small. Using equation (\ref{corr}), it is easy to see that the
basic equations (\ref{eq200}) and (\ref{general}), which determine the
optimal price and the hedging strategy, can be solved perturbatively in $\mu
_{n}$. For our purposes, we will only need the first correction in the
conditional drift, which we now derive. First note that equation (\ref{eq200}%
) can be rewritten as 
\begin{equation*}
\mathcal{C}=\left\langle \Theta \right\rangle -\sum_{n=0}^{N-1}\left\langle
\phi _{n}^{\ast }\mu _{n}\right\rangle .
\end{equation*}
In particular, to first order in $\mu _{n}$, one finds that 
\begin{equation}
\mathcal{C}=\left\langle \Theta \right\rangle -\sum_{n=0}^{N-1}\left\langle 
\frac{\Delta _{n+1}\mu _{n}}{D_{n}}\Theta \right\rangle .  \label{eq1000}
\end{equation}
One can also compute the first correction to the hedging strategy, by
combining equation (\ref{corr}) with the zero--th order results obtained in
the last section. One obtains then 
\begin{equation}
\phi _{0}^{\star }=\frac{1}{D_{0}}\left\langle \Delta _{1}\Theta
\right\rangle -\frac{1}{D_{0}}\sum_{n=1}^{N-1}\left\langle \frac{\Delta
_{n+1}\mu _{n}}{D_{n}}\Delta _{1}\Theta \right\rangle -\frac{1}{D_{0}}\mu
_{0}\left\langle \Theta \right\rangle .  \label{eq1001}
\end{equation}
In the above equation, we have only shown the optimal hedge for the first
time period $\phi _{0}^{\star }$. The explicit expression for the optimal
hedge for the subsequent time periods is also of a similar form, but it is
quite intricate and not very illuminating. More importantly, it is not
necessary to determine the correct hedging strategy. This important point
will be discussed, in a slightly more general context, in section \ref{other}.

\section{\label{model}A general model for weakly correlated processes}

Let us introduce a rather general model for correlated price increments. We
first define an auxiliary set of \textit{uncorrelated i.i.d. random variables%
} $\{\eta _{n}\}$, distributed according to an arbitrary probability
distribution\footnote{%
The following results can be generalized to the case where the distribution $%
P(\eta )$ explicitly depends on $n$.} $P(\eta )$. The joint distribution is
then given by 
\begin{equation*}
P(\{\eta _{n}\})=\prod_{n}P(\eta _{n}).
\end{equation*}
We will assume that $P(\eta )$ is symmetrical\footnote{%
In the sequel, we will actually only need that equation (\ref{eq600}) holds
for $q=0$,$1$.} -- \textit{i.e. }that 
\begin{equation}
\int \eta ^{2q+1}P(\eta )d\eta =0,  \label{eq600}
\end{equation}
and we will denote the second moment of $P$ by 
\begin{equation*}
D=\int \eta ^{2}P(\eta )d\eta .
\end{equation*}
We now construct the set of correlated price increments $\{\Delta _{n}\}$ by writing 
\begin{equation}
\Delta _{n}=\sum_{m}M_{nm}\eta _{m}+\mu ,  \label{eq400}
\end{equation}
with 
\begin{equation*}
M_{nm}=\delta _{n,m}+\frac{1}{2D}c_{n-m}.
\end{equation*}
The coefficients $c_{n}$ are assumed to satisfy 
\begin{equation*}
c_{0}=0\qquad \ \ \ \ \ \ \ \ \ \ \ \ \ c_{-k}=c_{k}.
\end{equation*}
In the sequel, we assume that both the $c_{k}$'s and $\mu $ are small, and
will work only to first order in both $\mu $ and$\ c$.

Let us show the significance of equation (\ref{eq400}), by first noting
that, since $\left\langle \eta _{n}\right\rangle =0$, we have that 
\begin{equation*}
\left\langle \Delta _{n}\right\rangle =\mu .
\end{equation*}
Therefore $\mu $ is nothing but the average \textit{unconditional} drift of
the price process. Moreover, to first order in $c$, the following holds 
\begin{equation*}
\left\langle (\Delta _{n}-\mu )(\Delta _{m}-\mu )\right\rangle
=\sum_{i,j}M_{ni}M_{mj}\left\langle \eta _{i}\eta _{j}\right\rangle =D\delta
_{n,m}+c_{n-m}.
\end{equation*}
The coefficients $D$ and $c$ denote then, respectively, the \textit{%
unconditional} variance and correlation of the price increments. Finally,
one has that, independently of $c_{n-m}$, and for all $p$, 
\begin{equation*}
\left\langle (\Delta _{n}-\mu )^{p}\right\rangle =\left\langle \eta
^{p}\right\rangle .
\end{equation*}

We note that the variance of the price differences over a time scale $n$,
given by 
\begin{equation*}
\sigma ^{2}(n)=\left\langle \left( \sum_{k=m+1}^{m+n}(\Delta _{k}-\mu
)\right) ^{2}\right\rangle \ ,\qquad (n\geq 1)
\end{equation*}
is related to the correlation coefficients $c_{k}$ as follows 
\begin{eqnarray}
\sigma ^{2}(n) &=&nD+\sum_{k=1}^{n-1}2(n-k)c_{k}  \notag \\
c_{n} &=&\frac{1}{2}\left( \sigma ^{2}(n+1)-2\sigma ^{2}(n)+\sigma
^{2}(n-1)\right) .  \label{eq33}
\end{eqnarray}
In the case where $c_{n}=0$, one finds $\sigma ^{2}(n)=nD$, whereas, in the
case of very short range anticorrelations ($c_{n}=-c\delta _{n,1}$), one has 
$\sigma ^{2}(n)=n(D-2c)+2c$. In this case, the global volatility $\sigma
^{2}(n)/n\rightarrow D-2c$ is lower than the short--range volatility $\sigma
^{2}\left( 1\right) =D$.

The probability distribution of the price increments can be written, to
first order in $\mu ,c$, as 
\begin{eqnarray}
&&\prod_{n}P\left(\Delta _{n}-\mu -\frac{1}{2D}\sum_{m}c_{n-m}\Delta _{m}\right)
\label{dist1} \\
&=&\left[ 1-\mu \sum_{n}\frac{\partial \ln P}{\partial \Delta _{n}}-\frac{1}{%
2D}\sum_{n,m}\frac{\partial \ln P}{\partial \Delta _{n}}c_{n-m}\Delta _{m}%
\right] \prod_{n}P(\Delta _{n}).  \notag
\end{eqnarray}
It is now easy to write the marginal distribution at time $n=0$, given all
previous returns $\Delta _{n}$ with $n\leq 0$ 
\begin{equation}
\left[ 1-\mu \sum_{n>0}\frac{\partial \ln P}{\partial \Delta _{n}}-\frac{1}{%
2D}\sum_{\max (n,m)>0}\frac{\partial \ln P}{\partial \Delta _{n}}%
c_{n-m}\Delta _{m}\right] \prod_{n>0}P(\Delta _{n}).  \label{condi}
\end{equation}
The equation above follows immediately from the general expression for the
probability distribution, up to a possible normalization factor. To show
that (\ref{condi}) is actually correctly normalized, we have only to note
the obvious fact that 
\begin{equation*}
\int d\Delta \,P(\Delta )\frac{\partial \ln P}{\partial \Delta }=0.
\end{equation*}
Moreover, using that 
\begin{eqnarray*}
\int d\Delta \,P(\Delta )\Delta \frac{\partial \ln P}{\partial \Delta } &=&-1
\\
\int d\Delta \,P(\Delta )\Delta ^{2}\frac{\partial \ln P}{\partial \Delta }
&=&0
\end{eqnarray*}
it is then easy to show that the \textit{conditional }drift and variance $%
\mu _{n}$ and $D_{n}$ are given, to first order in $\mu$ and $c$, by 
\begin{eqnarray*}
\mu _{n-1} &=&\mu +\sum_{m<n}c_{n-m}\left( \frac{1}{2D}\Delta _{m}-\frac{1}{2%
}\frac{\partial \ln P}{\partial \Delta _{m}}\right) \\
D_{n} &=&D.
\end{eqnarray*}

\section{\label{pricing}Derivative pricing with small correlations}

\subsection{A general formula}
Let us now apply the general perturbative formula (\ref{eq1000}) for the
price of the contract to the specific model of correlations described in the
previous section . One finds 
\begin{eqnarray*}
\mathcal{C} &=&\left\langle \Theta \right\rangle -\frac{1}{D}%
\sum_{n=1}^{N}\left\langle \mu _{n-1}\Delta _{n}\Theta \right\rangle \\
&=&\left\langle \Theta \right\rangle -\frac{\mu }{D}\sum_{n=1}^{N}\left%
\langle \Delta _{n}\Theta \right\rangle -\frac{1}{2D}\sum_{n=1}^{N}%
\sum_{m<n}c_{n-m}\left\langle \left( \frac{1}{D}\Delta _{m}-\frac{\partial
\ln P}{\partial \Delta _{m}}\right) \Delta _{n}\Theta \right\rangle .
\end{eqnarray*}
Now, the above averaging $\left\langle \cdots \right\rangle $ is performed
over the `correlated' historical price increment distribution, equation (\ref
{condi}). It is convenient to re--express all of the following formul\ae\ in
terms of an auxiliary probability distribution $\prod_{n>0}P\left( \Delta
_{n}\right) $, which describes an unbiased and uncorrelated process, and is
obtained by setting $c_{k}=\mu =0$. We will denote the corresponding
averages by $\langle ...\rangle ^{0}$, where the superscript indicates that
the conditional drift is now set to zero (in the case of a risk--free
Gaussian model, this auxiliary probability distribution is called the
equivalent martingale measure in the financial mathematics literature).
Equation (\ref{condi}) can then be reinterpreted as relating the two
averaging procedures $\left\langle \cdots \right\rangle $ and $\left\langle
\cdots \right\rangle ^{0}$ as follows 
\begin{equation*}
\left\langle \cdots \right\rangle =\left\langle \cdots \right\rangle^0
-\mu \sum_{n>0}\left\langle \cdots \frac{%
\partial \ln P}{\partial \Delta _{n}}\right\rangle ^{0}-\frac{1}{2D}%
\sum_{\max (n,m)>0}\left\langle \cdots \frac{\partial \ln P}{\partial \Delta
_{n}}c_{n-m}\Delta _{m}\right\rangle ^{0}.
\end{equation*}
After a some algebra we then find that 
\begin{equation}
\mathcal{C}=\left\langle \Theta \right\rangle ^{0}+\mu
\sum_{n=1}^{N}\left\langle F\left( \Delta _{n}\right) \Theta \right\rangle
^{0}+\frac{1}{2D}\sum_{n=1}^{N}\sum_{m<n}c_{n-m}\left\langle F(\Delta
_{n})\Delta _{m}\Theta \right\rangle ^{0},  \label{price}
\end{equation}
where 
\begin{equation*}
F(\Delta )=-\frac{\partial \ln P}{\partial \Delta }-\frac{1}{D}\Delta .
\end{equation*}
The above equation is the central result of this paper, that we now comment
in various limits.

Let us first consider the purely Gaussian case, where $\ln P(\Delta
)=-\Delta ^{2}/2D-1/2\ln (2\pi D)$. In this situation, it is easy to check
that $F(\Delta )=0$. Therefore, all corrections brought about by the drift
and correlations strictly vanish, and the price can be calculated by
assuming that the process is correlation--free. As shown in 
the Appendix, this is indeed expected in a continuous time framework.
In other words, the price of
the option is given by the Black--Scholes price, with a volatility given by
the small scale volatility $D$ (that corresponds to the hedging time scale),
and {\it not} with the volatility corresponding to the terminal distribution, $\sigma
^{2}(N)/N$. As we have already shown, when price increments are
anticorrelated, $D$ is larger than $\sigma ^{2}(N)/N$, and the price of the
contract is higher than the objective average pay--off. This is due to the
fact that the hedging strategy is on average losing money because of the
anticorrelations: a rise of the price of the underlying leads to an increase
of the hedge, which is followed (on average) by a drop of the price.

For an uncorrelated non--Gaussian process, the effect of a non--zero drift $%
\mu $ does not vanish, a situation already discussed in details in \cite{us}%
. The same is true with correlations: the correction no longer disappears
and one cannot price the option by considering a process with the same
statistics of price increments but no correlations. Note that the correction
term involves all $\Delta _{m}$ with $m<0$. This means, as expected a
priori, that, in the presence of correlations, the price of the option does
not only depend on the current price of the underlying, but also on all past
price increments.

\subsection{The case of short range correlations}

Let us illustrate the general formula in the case of very short range
anticorrelations $c_{n}=-c\delta _{n,1}$. We will furthermore consider the
simple case of path--\textit{independent} pay--offs $\Theta $, which are
functions only on the terminal price $x_{N}$ of the underlying.

The calculation of the price will therefore involve the following quantities 
\begin{equation*}
I_{1}=\langle F\left( \Delta _{n}\right) |x_{N}\rangle ^{0}\ \ \ \ \ \ \ \ \
\ \ \ \ \ \ \ \ \ \ I_{2}=\langle \Delta _{n}F\left( \Delta _{m}\right)
|x_{N}\rangle ^{0},
\end{equation*}
where the notation $\langle ...|x_{N}\rangle ^{0}$ means that we are
conditioning on a given value of the terminal price $x_{N}$. The computation
of the two quantities $I_{1},I_{2}$ is easily done in momentum space, with
the aid the following formulae 
\begin{eqnarray*}
\int d\Delta \;\Delta P(\Delta) e^{iz\Delta } &=&iDz-\frac{i}{6}D^{2}\left( 3+\kappa
\right) z^{3}+o\left( z^{5}\right)  \\
\int d\Delta \;F\left( \Delta \right)P(\Delta) \,e^{iz\Delta } &=&\frac{i}{6}\kappa
D\,z^{3}+o\left( z^{5}\right) ,
\end{eqnarray*}
where $\kappa$ is the kurtosis of the distribution of elementary increments. 
The result reads, for large $N$,
\begin{equation*}
I_{1}=-\frac{\kappa }{N\sqrt{DN}}p_{1}\;\ \ \ \ \ \ \ \ \ \ \ \ \ \ \ \ \
I_{2}=-\frac{4\kappa }{N^{2}}p_{2},
\end{equation*}
where $p_{1}$ and $p_{2}$ are, respectively, the skewness and kurtosis
polynomials 
\begin{equation*}
p_{1}=\frac{1}{6}\left( \widetilde{x}^{3}-3\widetilde{x}\right) \;\ \ \ \ \
\ \ \ \ \ \ \ \ \ \ \ p_{2}=\frac{1}{24}\left( \widetilde{x}^{4}-6\widetilde{%
x}^{2}+3\right) 
\end{equation*}
and where we have consider the natural scaling $\widetilde{x}^{2}=\left(
x_{N}-x_{0}\right) ^{2}/DN$ : $\widetilde{x}$ is the moneyness of the option
counted in standard deviations. The general equation (\ref{price}) then
takes the form $\mathcal{C}=\langle \widetilde{\Theta}\rangle ^{0}$, with a
modified pay--off function 
\begin{equation*}
\widetilde{\Theta}(x_{N})=\Theta (x_{N})\left[ 1-\frac{\kappa }{\sqrt{DN}}\left(
\mu -\frac{c\Delta _{0}}{2DN}\right) p_{1}+\frac{2c\kappa }{DN}p_{2}\right] ,
\end{equation*}
where the explicit dependence on the last past increment $\Delta _{0}$
appears. Note that the correction terms vanish for at the money options ($%
x_{N}=x_{0}$). Let us concentrate on the last term in the above equation. It
represents an \textit{increase} in the global kurtosis $\kappa /N$ of the
`risk neutral' distribution that must be used to price the option:
\begin{equation}
\frac{\kappa }{N}\rightarrow \frac{\kappa }{N}\left( 1+\frac{2c}{D}\right) .
\label{res20}
\end{equation}
This is means that the `implied' kurtosis is increased, in percent terms, as much as
the global variance is decreased from the local one by correlation effects.
Therefore the volatility `smile', which is proportional to the kurtosis, is
enhanced by the presence of anti--correlations, and reduced by the presence of
correlations. Note that the above model neglects all volatility fluctuations,
which lead to a terminal kurtosis decaying much more slowly than $\kappa/N$ (see \cite{us}).
The above calculations could in principle be extended to deal with this effect, which
is crucial for financial applications.

\subsection{An alternative model of correlations}

Let us comment more on the above result. To this end, let us briefly discuss
an other possible model of correlated price increments. One could consider,
instead of the probability distribution (\ref{dist1}), the following
distribution for price increments (we set for simplicity $\mu =0$) 
\begin{equation}
\left[ 1+\frac{1}{2D^{2}}\sum_{n,m}c_{n-m}\Delta _{n}\Delta _{m}\right]
\prod_{n}P(\Delta _{n}).  \label{dist2}
\end{equation}
The above measure induces statistics similar to (\ref{dist1}), and in
particular one still has that $\left\langle \Delta _{n}\Delta
_{m}\right\rangle =D\delta _{n,m}+c_{n-m}$, and that $\left\langle \Delta
_{n}^{p}\right\rangle =\left\langle \eta ^{p}\right\rangle $. On the other
hand, it is not hard to show, following an analysis similar to the one in
section \ref{general}, that, to first order in the correlations, the price
of an option is \textit{unchanged, }regardless of the basic day--to--day
probability distribution $P\left( \eta \right) $. How can one decide which
model more accurately describes a specific time--series of price increments?
To answer this question, we first note that (\ref{dist1}) and (\ref{dist2})
coincide when $P\left( \eta \right) $ is Gaussian. It must then be that the
deviations involve the higher moments of $P$. Let us denote with $\lambda
_{i}$ the $i$--th cumulant of $P$ (in particular $\lambda _{2}=D$ and $%
\lambda _{4}=D^{2}\kappa $, with $\kappa $ the kurtosis), and with $\lambda
_{i}\left( n\right) $ the $i$--th cumulant of the return $\Delta _{1}+\cdots
+\Delta _{n}$ of the stock in $n$ time periods. If the price increments
where independent, then $\lambda _{i}\left( n\right) =n\lambda _{i}$. In the
presence of correlations, this equation is not valid any more, but, if the
correlations are short--range, we can define global cumulants as 
\begin{equation*}
\widetilde{\lambda }_{i}=\frac{1}{n}\lambda _{i}\left( n\right) .\;\ \ \ \ \
\ \ \ \ \ \ \ \ \ \ \ \ \ \ \ \left( n\rightarrow \infty \right) 
\end{equation*}
To be definite, let us consider the case already discussed of short--range
anticorrelations $c_{1}=-c$. We have already shown that $\lambda _{2}\left(
n\right) =n\left( D-2c\right) +2c$ and therefore that 
\begin{equation*}
\widetilde{\lambda }_{2}=D\left( 1-\frac{2c}{D}\right) .
\end{equation*}
It is then not hard to show that, if one considers the model (\ref{dist1}),
then the global forth moment is given by 
\begin{equation*}
\widetilde{\lambda }_{4}=D^{2}\kappa \left( 1-\frac{4c}{D}\right) .
\end{equation*}
On the other hand, if one uses (\ref{dist2}) one finds that 
\begin{equation*}
\widetilde{\lambda }_{4}=D^{2}\kappa \left( 1-\frac{8c}{D}\right) .
\end{equation*}
In both cases, the global historical kurtosis is reduced by anti--correlation effects,
but the amount varies in the two models. Recalling that $2c/D\simeq 0.10$,
we see that the two models differ by $10\%$ in the measurement of the global
kurtosis $\widetilde{\lambda }_{4}/D^{2}\kappa $.

\subsection{The case of Delta-hedging} 

Le us conclude this section by discussing other possible hedging schemes.
Instead of choosing the optimal hedge that minimizes the variance, one can
follow the (sub--optimal) Black--Scholes $\Delta $--hedge. As emphasized in 
\cite{us}, this leads in general to a negligible risk increase, and has the
advantage of removing the effect of the drift $\mu $ on the price, and
therefore reduces the risk associated to change of long time trends, encoded
in $\mu $. One might therefore expect that $\Delta $--hedging also allows
one to get rid of the correlations. However, the following calculation shows
that this is not the case.

Up to first order in $c$, it is sufficient to consider that the $\Delta $%
--hedge is given by 
\begin{equation*}
\phi _{BS,n}=\frac{\partial }{\partial x_{n}}\langle \Theta \rangle
_{n}^{0}=-\langle \frac{\partial \ln P}{\partial \Delta }(\Delta
_{n+1})\Theta \rangle _{n}^{0}.
\end{equation*}
The price of the contract is then given by 
\begin{eqnarray*}
\mathcal{C} &=&\left\langle \Theta \right\rangle
-\sum_{n=0}^{N-1}\left\langle \mu _{n}\phi _{BS,n}\right\rangle  \\
&=&\left\langle \Theta \right\rangle +\sum_{n=0}^{N-1}\left\langle \mu _{n}%
\frac{\partial \ln P}{\partial \Delta _{n+1}}\Theta \right\rangle  \\
&=&\langle \Theta \rangle ^{0}+\frac{1}{2}\sum_{n=1}^{N}\sum_{m<n}c_{m-n}%
\langle F(\Delta _{n})\frac{\partial \ln P}{\partial \Delta _{m}}\Theta
\rangle ^{0}.
\end{eqnarray*}
Note that, in this case, the drift $\mu $ indeed disappears to first order
even if $F(\Delta )$ is not zero; the correlations, on the other hand, are
still present within a $\Delta $--hedging scheme.

\section{\label{other}Other optimization schemes}

Let us now consider a slightly different problem, where the price of the
option is fixed by the market to a certain value $\mathcal{C}_{M}$. The
option trader knows that price increments have a non--zero conditional drift
and wishes to buy and hedge the option in order to optimize a certain risk
adjusted return. The wealth balance associated to selling the option now
reads 
\begin{equation*}
W=\mathcal{C}_{M}+H-\Theta ,
\end{equation*}
so that the expected return is equal to 
\begin{equation*}
M=\mathcal{C}_{M}+\langle H-\Theta \rangle =\mathcal{C}_{M}-\mathcal{C}
\end{equation*}
and the expected risk is 
\begin{equation*}
\mathcal{R}^{2}=\left\langle W^{2}\right\rangle -M^{2}.
\end{equation*}
We will assume that the trader wants to maximize a certain function $G$ of $%
M $ and $\mathcal{R}$, for example the Sharpe ratio $G_{1}=M/\mathcal{R}$,
or a certain risk corrected return $G_{2}=\lambda M-\mathcal{R}$.

Consider a small variation $\delta \phi _{n}$ of the strategy $\phi _{n}$.
The corresponding changes in $M$ and $\mathcal{R}$ read 
\begin{equation*}
\frac{\delta M}{\delta \phi _{n}}=\mu _{n}\qquad \mathcal{R}\frac{\delta 
\mathcal{R}}{\delta \phi _{n}}=\langle \Delta _{n+1}(H-\Theta )\rangle _{n}+%
\mathcal{C}\,\mu _{n}.
\end{equation*}
We may now extremize $G$ with respect to the hedging strategy and obtain the
following equation 
\begin{equation*}
\frac{\delta M}{\delta \phi _{n}}\frac{\partial G}{\partial M}+\frac{\delta 
\mathcal{R}}{\delta \phi _{n}}\frac{\partial G}{\partial \mathcal{R}}=0,
\end{equation*}
or 
\begin{equation}
\langle \Delta _{n+1}(\Theta -H)\rangle _{n}=\mu _{n}\mathcal{C}+\mu _{n}%
\mathcal{G\;\ \ \ \ \ \ \ \ G}=\mathcal{R}\frac{\partial _{M}G}{\partial _{%
\mathcal{R}}G}.  \label{eq2000}
\end{equation}
In the two examples considered before, one has that 
\begin{equation*}
\mathcal{G}_{1}=-\frac{\mathcal{R}^{2}}{M}\qquad \ \ \ \ \ \mathcal{G}%
_{2}=-\lambda \mathcal{R}.
\end{equation*}
Equation (\ref{eq2000}) is a generalization of equation (\ref{general}), and
is again well suited for a perturbation in $\mu $. We will work as always to
first order in $\mu $, and we will call $\widetilde{\phi}$ the solution to (\ref
{eq2000}) with the function$\mathcal{\ G}$ set to zero -- that is, the
solution to the problem solved in section \ref{pert}). It is then easy to
see, using equation (\ref{corr}), that the optimal strategy is given by 
\begin{equation}
\phi _{n}^{\ast }=\widetilde{\phi}_{n}-\frac{\mu _{n}}{D_{n}}\left. \mathcal{G}%
\right| _{\widetilde{\phi}}.  \label{eq3000}
\end{equation}

We conclude this section by showing that the more general formalism
developed above actually has some interesting consequences for the original
problem of option pricing and hedging which was considered in the rest of
the paper. Let us in particular consider, as we have done in the previous
sections, a specific contract sold at time $0$ at the optimal price $%
\mathcal{C}_{0}$ and with optimized hedging strategy $\phi _{n,0}^{\star }\,$%
(we will explicitly insert the time--index for the date of \textit{sale} of
the contract in the last part of this section). More generally, we may sell
the same contract at some other time $P$ at a price $\mathcal{C}_{P}$ with a
hedging strategy $\phi _{n,P}^{\star }$ (with $n\geq P$). It is a natural
question to ask what is the relation of the various hedging strategies $\phi
_{n,P}^{\star }$ for \textit{fixed} time $n$, as we vary the sale time of
the contract $P$. In particular, one may ask if the optimal hedging strategy 
$\phi _{n,P}^{\star }$ at time $n$ should be equal to the optimal hedge $%
\phi _{n,n}^{\star }$ which one is to adopt if one is selling the contract
at time $n$. This question can be easily answered in the framework of the
present section, by noting that, at time $n$, the option traders wealth
balance is given by the original price of the contract $\mathcal{C}_{P}$
plus the returns of the hedging strategy up to time $n$%
\begin{equation*}
\mathcal{W}_{n}=\mathcal{C}_{P}+\phi _{P,P}^{\star }\Delta _{P+1}+\cdots
+\phi _{n-1,P}^{\star }\Delta _{n}.
\end{equation*}
From the point of view of the trader, the above quantity acts as an
effective market price $\mathcal{C}_{M}$, and the optimization problem after
time $n$ is of the type described at the beginning of this section.
Recalling that the trader is trying to minimize $\left\langle
W^{2}\right\rangle $, we deduce that the function $G$ is given by $M^{2}+%
\mathcal{R}^{2}$, and therefore that $\mathcal{G}=M=\mathcal{C}_{M}-\mathcal{%
C}=\mathcal{W}_{n}-\mathcal{C}_{n}$. We can then use equation (\ref{eq3000})
to deduce that 
\begin{equation*}
\phi _{n,P}^{\star }=\phi _{n,n}^{\star }-\frac{\mu _{n}}{D_{n}}\left( 
\mathcal{W}_{n}-\mathcal{C}_{n}\right).
\end{equation*}
The above equation has a clear interpretation. In a risk--free model, the
quantity $\left( \mathcal{W}_{n}-\mathcal{C}_{n}\right) $ vanishes, since we
can, by assumption, perfectly reproduce the price--process of an option by
hedging correctly. Therefore the correct hedge is independent of the time of
sale of the contract. On the other hand, in a model with non--zero risk, one
needs to correct the hedging strategy whenever the past hedge has not
completely compensated the price change of the option. In particular, if the
hedging strategy has made too much money and $\mathcal{W}_{n}>\mathcal{C}_{n}$%, one 
has to decrease the amount of stock in the portfolio whenever the
conditional drift $\mu _{n}$ is positive, and increase it otherwise. A similar 
result was obtained in \cite{us}, section 3.4; it was however noted there that this
history dependent strategy, although reducing the variance, actually increases the
probabilty of large losses.

\section{\label{conc}Summary and conclusion}

We have considered the problem of option pricing and hedging when stock
returns are correlated in time. Using a variance minimization framework, we
have obtained a general formula, valid for weakly correlated non--Gaussian
processes. We have shown that in the limit of Gaussian price increments, the
correlations are irrelevant, in the sense that the Black--Scholes formula
holds with volatility that of the price increments on the time--scale of the
re--hedging. For non--Gaussian processes, however, further non trivial
corrections to the `smile' effect are brought about by the correlations,
even when the hedge is the Black--Scholes $\Delta $--hedge. The kurtosis to
be used in option--pricing is neither the local one, nor the global one, but
some effective kurtosis which depends on the strength of the correlations.
The above formalism can be extended to treat the case where the distribution
of price increments depends explicitly on time, as would be needed to treat
the case of a time dependent volatility. Finally, we have given formula for
the optimal hedge when the expected return associated to options is non zero.

\section*{Appendix: Correlations within the Black--Scholes framework}

The presence of correlations in the continuous time limit can be modelled as
a history dependent drift term in the stochastic differential equation for
the price (or log-price process). More precisely, the dynamical evolution of
the price reads:
\begin{equation}
dX(t) = \mu(t) dt + \sigma_0 dW(t)
\end{equation}
where the drift $\mu(t)$ depends on the whole past history and $dW$ is a
Brownian noise of unit variance per unit time. A simple model of correlations is to 
assume that $\mu(t)$ is constructed from the past values of $dW(t')$ as
an exponential moving average:
\begin{equation}
\mu(t)= \epsilon \Gamma \int_{-\infty}^t dW(t') e^{-\Gamma(t-t')}.
\end{equation}
When $\epsilon > 0$, this corresponds to correlations in price increments
extending over a time scale $\Gamma^{-1}$. Conversely, $\epsilon < 0$ corresponds
to anticorrelations.
It is easy to see that in this model, the time dependent square volatility $\sigma^2(t)$ is
given by:
\begin{equation}
\sigma^2(t)=\frac{1}{t} \int_0^{t} dt' \left(\sigma_0 + \epsilon (1-e^{-\Gamma{t-t'}})
\right)^2,
\end{equation}
that interpolates between the short term volatility $\sigma_0$ and the long term volatility
$|\sigma_0+\epsilon|$. 

In the Black--Scholes formalism, however, the drift is totally absent from the price
and hedge of an option. This is intimately related to the fact that the Ito formula
does not involve the drift $\mu(t)$ \cite{Baxter}. Hence, it is immediate that in this case, correlations
are irrelevant and only the short term volatility $\sigma_0$ is needed to price the option.
This is what we find within our general formalism in the case of Gaussian increments.

Note that the difference between the short term volatility and long term volatility is
particularly striking in the case of an Ornstein-Uhlenbeck (mean reverting) process for
which $\mu(t)=-K X(t)$. In this case, the long term volatility tends to zero: $X(t)$
has bounded fluctuations, even when $t \to \infty$. Correspondingly, the average pay-off
over the objective probability distribution is finite for $t \to \infty$, whereas the 
option price, given by the standard Black--Scholes formula with a volatility $\sigma_0$,
tends to infinity. The difference between the two comes from the hedging strategy that
loses, because of the anticorrelations, an infinite amount of money in the limit 
$t \to \infty$.

\end{document}